\documentclass[a4paper,10pt]{article}

\usepackage[a4paper, portrait]{geometry}
\usepackage[applemac]{inputenc}
\usepackage{amsmath}
\usepackage{slashed}
\usepackage{amscd}\usepackage{amstext}\usepackage{amsbsy}\usepackage{amsopn}\usepackage{amsthm}\usepackage{amsxtra}\usepackage{upref}\usepackage{amsfonts}
\usepackage{amssymb}\usepackage{euscript}
\usepackage[]{latexsym}\usepackage{graphicx}\usepackage{color}\usepackage[all]{xy}
\usepackage[american]{babel}
\usepackage{slashed}\usepackage{physics}
\usepackage{tikz}
\usepackage{tikz-3dplot}
\usepackage{mdframed}
\usepackage{todonotes}
\usetikzlibrary{patterns}
\usepackage{hyperref}
\usepackage{pgfplots}
\usepackage{authblk} 
\usepackage{wrapfig}
\usepackage{subcaption}
\usepackage{cleveref}

\usepackage{caption}
\newmdtheoremenv{framedtheorem}{Theorem}

\title{Inverse algorithm and triple point diagrams}
\author[1]{Valdo Tatitscheff \thanks{valdo.tatitscheff@normalesup.org}} 
\affil[1]{\footnotesize{IRMA, UMR 7501, Universit\'e de Strasbourg et CNRS\\ 
		7 rue Ren\'e Descartes 67000 Strasbourg, France}}

\date{}

\begin{document}

\maketitle

\vspace{-1cm}

\abstract{Dimer models (also known as brane tilings) are special bipartite graphs on a torus $\mathbb{T}^2$. They encode the structure of the 4d $\mathcal{N}=1$ worldvolume theories of D3 branes probing toric affine Calabi-Yau singularities. Constructing dimer models from a singularity can in principle be done via the so-called inverse algorithm, however it is hard to implement in practice. We discuss how combinatorial objects called triple point diagrams systematize the inverse algorithm, and show how they can be used to construct dimer models satisfying some symmetry or containing particular substructures. We present the construction of the Octagon dimer model which satisfies both types of constraints. Eventually we present a new criterion concerning possible implementations of symmetries in dimer models, in order to illustrate how the use of triple point diagrams could strengthen such statements.}


\section{Dimer models}

\begin{figure}[h!]
	\centering
	\begin{subfigure}{0.27\textwidth}
		\centering
		\includegraphics[width=0.7\linewidth]{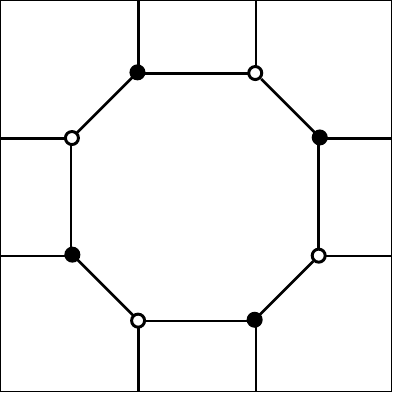} 
	\end{subfigure}
	\begin{subfigure}{0.27\textwidth}
		\centering
		\includegraphics[width=0.7\linewidth]{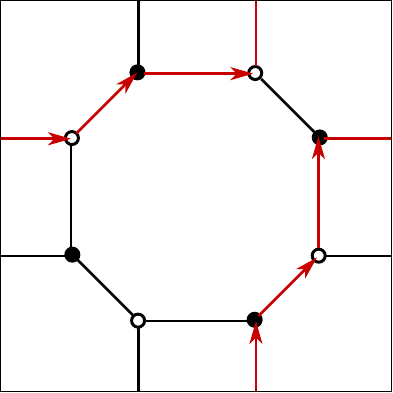} 
	\end{subfigure}
	\begin{subfigure}{0.27\textwidth}
		\centering
		\includegraphics[width=0.7\linewidth]{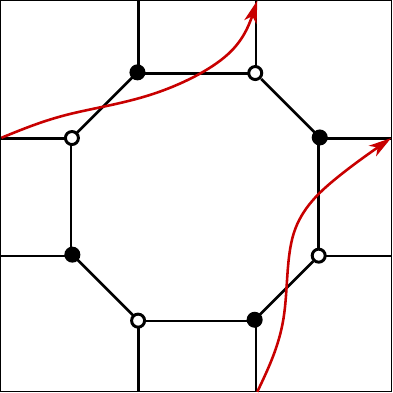} 
	\end{subfigure}
	
	\caption{A dimer model, a zig-zag path and a zig-zag strand.}
	\label{Fig:DMandZZP}
\end{figure}

For our purposes a \textit{dimer model} (or \textit{brane tiling}) is a bipartite graph $\Gamma$ embedded in a topological $2$-torus $\mathbb{T}^2$ in such a way that each connected component of $\mathbb{T}^2\backslash\Gamma$ is simply connected, and considered up to isotopy. An example of dimer model is given on the left of \Cref{Fig:DMandZZP}, where the square is a fundamental cell for $\mathbb{T}^2$. Fixing an orientation of the torus, it induces at each vertex a cyclic orientation of the edges incident to it, hence there is a globally well-defined notion of `left' and `right' at each vertex of $\Gamma$.

\vspace{-0.25cm}

A \textit{zig-zag path} (ZZP) on $\Gamma$ is an oriented path of edges of $\Gamma$ which turns maximally right at each black vertex, and maximally left at each white one. One can equivalently represent zig-zag paths as strands crossing edges at their middle. A zig-zag path and the corresponding strand are displayed respectively in the middle and on the right of \Cref{Fig:DMandZZP}. Note that this ZZP has a non-trivial winding around the torus: in the basis of homology induced by our choice of fundamental cell, its winding is $(1,1)$.

A dimer model is said to be \textit{consistent} if 1) there is no topologically trivial ZZP 2) no lift of ZZP in the universal cover $\mathbb{R}^2$ of the torus self-intersects 3) any two distinct lifts of ZZPs in the universal cover of the torus never intersect more than once in the same direction. These technical conditions are equivalent to the existence of (possibly degenerate) isoradial embeddings of $\Gamma$ in $\mathbb{T}^2$ \cite{Hanany:2005ss,Broomhead:2012wzy,1012.5449}. In what follows we always assume dimer models to be consistent.

Dimer models have been introduced in string theory as a tool to study the worldvolume quantum field theories (QFTs) of D3-branes probing affine toric Calabi-Yau threefolds (CY3) in type IIB string theory (see \cite{Yamazaki:2008bt} and references therein). These setups have proved very fruitful in extending the gauge/gravity correspondence to $\mathcal{N}=1$ conformal field theories and even non-conformal QFTs. A brane tiling encodes the structure of a quiver $\mathcal{N}=1$ QFT as follows: each face is an $\mathrm{SU}(N)$ gauge group, each edge is a matter field in the bifundamental representation of the faces of which it is the boundary, and each vertex encodes a monomial of the superpotential. Orientifolds can be implemented on dimer models as involutions of the bipartite map. They grant access to orthogonal and symplectic gauge groups, as well as matter in rank-$2$ tensor representations.

Each affine toric CY3 is in one-to-one correspondence with a class of lattice convex polygons (also called \textit{toric diagrams}) up to $\mathrm{SL}_2(\mathbb{Z})$, hence dimer models are also associated with lattice convex polygons up to $\mathrm{SL}_2(\mathbb{Z})$. Since each edge of a dimer model belongs exactly to two ZZPs going in opposite directions, the sum of windings of all ZZPs in a brane tiling is always zero. Ordering the windings according to their slope, one builds the sides of a convex lattice polygon. The choice of a fundamental cell for $\mathbb{T}^2$ translates as the action of $\mathrm{SL}_2(\mathbb{Z})$ on the corresponding polygon. In fact the lattice to which the vertices of the polygon belong can be naturally identified with the integral first homology group of the torus $H_1(\mathbb{T}^2,\mathbb{Z})\simeq\mathbb{Z}^2$. 

There are in general many dimer models corresponding to the same affine toric CY3. Going from a dimer model to the class of lattice polygons up to $\mathrm{SL}_2(\mathbb{Z})$ as we explained before is the \textit{forward algorithm} and it is in general not injective; going in the opposite direction and building consistent dimer models from a class of toric diagrams can be done via the \textit{inverse algorithm} which is one-to-many in general \cite{Feng:2005gw}. Even if techniques have been developed to make the inverse algorithm more efficient \cite{Hanany:2005ss,Feng:2005gw}, it remains a set of instructions which has to be carried out case-by-case with a deft hand. Below we explain how Thurston's triple point diagrams \cite{2004math......5482T} help systematizing the inverse algorithm and even allow the construction of dimer models satisfying combinatorial constraints. 

There are interesting combinatorial transformations of consistent dimer models, as the so-called \textit{spider moves} and the closely related \textit{urban renewals}. From a physics perspective, they encode Seiberg dualities of the corresponding $\mathcal{N}=1$ theories \cite{Franco:2005rj}. Mathematically, they are an instance of quiver mutations and link dimer models to cluster algebras and varieties. In particular, each class of consistent dimer models up to spider moves defines an algebraic integrable system whose phase space is a cluster $\mathcal{X}$-variety \cite{Goncharov:2011hp}. Not uncorrelated to this is the fact that dimer models encode double Bruhat cells in the affine Poisson-Lie groups $\widehat{\mathrm{PGL}}(N)$ \cite{Fock:2014ifa}.

Dimer models also appear in other fields of mathematics and physics -- for instance crystal melting and topological strings \cite{Iqbal:2003ds}.

\section{Triple points diagrams and fast-inverse algorithm}

Triple point diagrams have been introduced by D. Thurston in \cite{2004math......5482T}. A triple point diagram is a collection of oriented dimension-one manifolds with boundary, mapped smoothly into a disk (the image of a connected component is called a \textit{strand}) such that 1) exactly three strands cross at each intersection point 2) the endpoints of strands are distinct points on the boundary of the disk and no other point is mapped to the boundary 3) the orientations of the strands induce consistent orientations on the complementary regions. Triple point diagrams are considered up to isotopy. Theorem 3 of \cite{2004math......5482T} states that given a disk with $2n$ points on the boundary labeled `in' or `out' and such that the labels alternate as one goes around the boundary, any matching (bijection) between in and out points can be realized by a triple point diagram.

\begin{figure}[h!]
	\centering
	\begin{subfigure}{0.23\textwidth}
		\centering
		\includegraphics[width=0.9\linewidth]{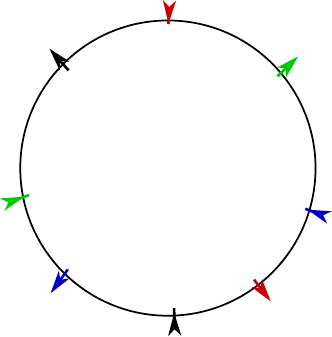} 
	\end{subfigure}
	\begin{subfigure}{0.23\textwidth}
		\centering
		\includegraphics[width=0.9\linewidth]{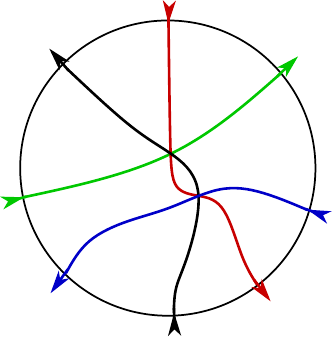} 
	\end{subfigure}
	\begin{subfigure}{0.23\textwidth}
		\centering
		\includegraphics[width=0.9\linewidth]{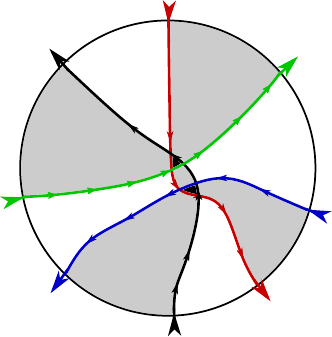} 
	\end{subfigure}
	\begin{subfigure}{0.23\textwidth}
		\centering
		\includegraphics[width=0.9\linewidth]{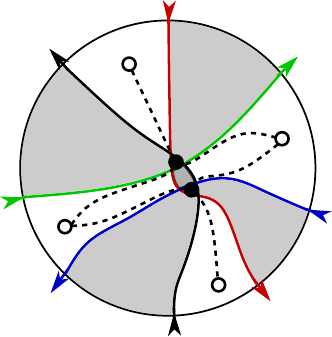} 
	\end{subfigure}
	\caption{From boundary data to bipartite graphs through triple crossing diagrams.}
	\label{Fig:Triplecrossings}
\end{figure}

\vspace{-0.25cm}

On the left of \Cref{Fig:Triplecrossings} such a boundary data on a disk is shown -- the pairing is displayed in colors. From it one easily constructs a triple point diagram. Deform the latter so that each crossing is replaced by a small triangle with counterclockwise oriented sides; one sees then that each connected component of the complement of the diagram in the disk has its boundary either oriented clockwise (white faces) or counterclockwise (black faces) or the orientation of edges along the boundary alternates (grey faces). Putting a black vertex in each black face, a white vertex in each white face and connecting vertices according to adjacency yields a bipartite graph on the disk. The so-called $2\leftrightarrow2$ moves of triple point diagrams correspond to spider moves of the graph.

Triple crossing diagrams can be used to systematically construct consistent dimer models from lattice polygons, as reviewed in \cite{Goncharov:2011hp}. One considers the simple outgoing normal vectors to the sides of a convex lattice polygon and for each of them one draws a straight line on a torus with winding the coordinates of the vector, such that the induced `in' and `out' points on the boundary alternate (it is always possible to do so \cite{Goncharov:2011hp}). Keeping only the boundary data together with the pairing between `in' and `out' points, forgetting momentarily the identifications between opposite sides of the square, one is left with a topological disk, boundary data and a pairing. Thus, one can construct a triple point diagram realizing the pairing and such that the number of triple crossings is minimal among such diagrams, deform it as before, and construct the corresponding bipartite graph. In the end we are left with a consistent dimer model on a torus that corresponds to the lattice polygon we started with. 

\begin{figure}[h!]
	\centering
	\begin{subfigure}{0.23\textwidth}
		\centering
		\includegraphics[width=0.8\linewidth]{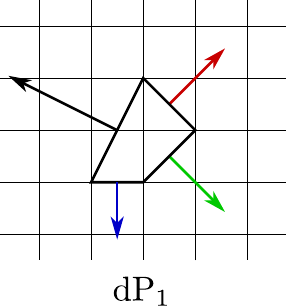} 
	\end{subfigure}
	\begin{subfigure}{0.23\textwidth}
		\centering
		\includegraphics[width=0.9\linewidth]{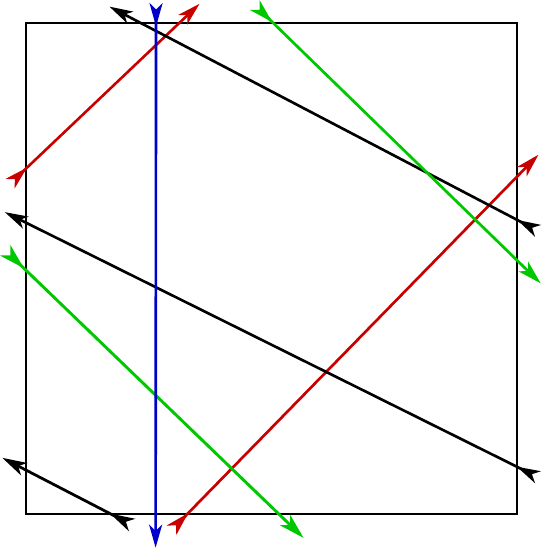} 
	\end{subfigure}
	\begin{subfigure}{0.23\textwidth}
		\centering
		\includegraphics[width=0.9\linewidth]{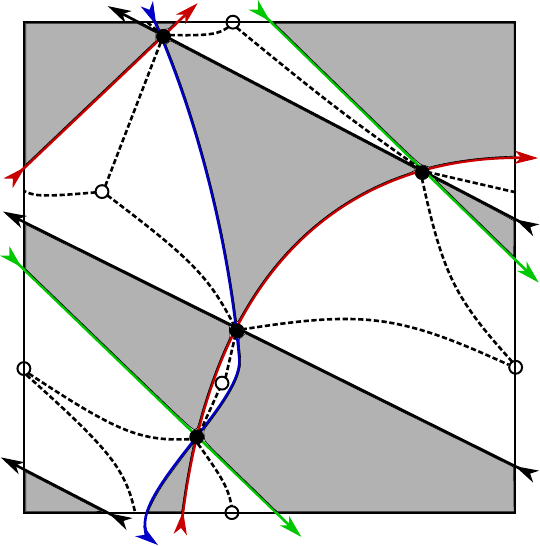} 
	\end{subfigure}
	\begin{subfigure}{0.23\textwidth}
		\centering
		\includegraphics[width=0.84\linewidth]{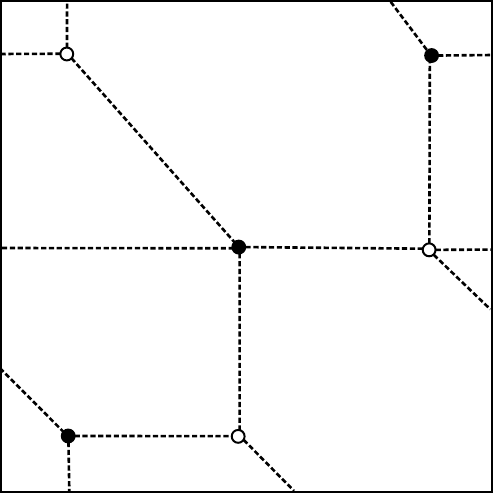} 
	\end{subfigure}
	\caption{From a toric diagram to a consistent dimer model through triple crossing diagrams.}
	\label{Fig:dp1}
\end{figure}

An example for a polygon dubbed $\mathrm{dP}_1$ is presented in \Cref{Fig:dp1}. On the right of the same figure is the dimer model we obtained. It famously corresponds to $\mathrm{dP}_1$ \cite{Franco:2005rj}.

\vspace{-0.25cm}

\section{The Octagon dimer model}

Supersymmetry (SUSY) has been introduced in the 80's in order to solve issues inherent to the Standard Model of particle physics, and grants a lot of control on the dynamics of the QFTs enjoying this property. However it is notoriously difficult to implement SUSY in a phenomenologically satisfactory way\footnote{many issues appear as one tries to explain how SUSY breaks spontaneously at low energies.}. Gauge theories in which SUSY is broken by dynamical effects are essential in many of these putative efforts; they are called Dynamical Supersymmetry Breaking (DSB) models. Only a few classes of explicit DSB models are known explicitly.

The physical motivation for the Octagon dimer model is to embed DSB models in brane tilings, for if it is possible it provides a natural UV completion of these DSB field theories. For a QFT to be implementable in a dimer model it must contain matter fields in tensor representations of rank at most $2$ of the gauge groups, which is the case for the so-called \textit{uncalculable $\mathrm{SU}(5)$} and the \textit{$3-2$} DSB models.  Orientifolds are needed in order to construct dimer models with rank-$2$ tensor matter.

While instances of brane tilings hosting DSB models are known since long \cite{Franco:2007ii}, it was shown more recently \cite{Buratti:2018onj,Argurio:2019eqb} that the presence of $\mathcal{N}=2$ fractional branes at the toric affine CY3 actually spoils the presence of the DSB vacuum. There are $\mathcal{N}=2$ fractional branes whenever the CY3 singularity is not-isolated, or equivalently when the corresponding toric diagrams are convex, but not strictly convex.

\begin{figure}[h!]
	\centering
	\begin{subfigure}{0.3\textwidth}
		\centering
		\includegraphics[width=0.7\linewidth]{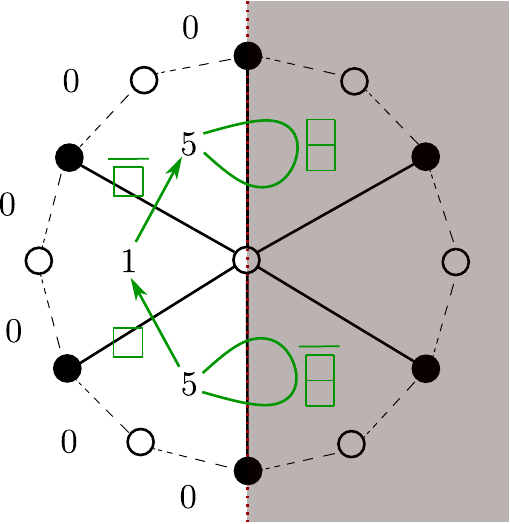} 
	\end{subfigure}
	\begin{subfigure}{0.3\textwidth}
		\centering
		\includegraphics[width=0.7\linewidth]{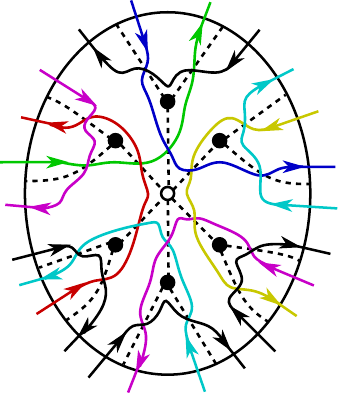} 
	\end{subfigure}
	\begin{subfigure}{0.3\textwidth}
		\centering
		\includegraphics[width=0.7\linewidth]{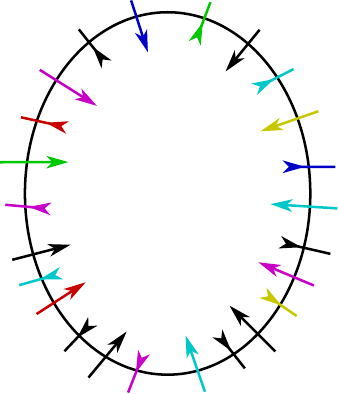} 
	\end{subfigure}
	\caption{The hexagonal cluster hosting the twin $\mathrm{SU}(5)$ model.}
	\label{Fig:hexagonalcluster}
\end{figure}

It was shown in \cite{Argurio:2020npm} that all possible substructures in dimer models leading to the $\mathrm{SU}(5)$ model imply the existence of $\mathcal{N}=2$ fractional branes, but the one shown on the left of \Cref{Fig:hexagonalcluster}. That led naturally to the question of whether it is possible at all to construct a dimer model corresponding to a strictly convex lattice polygon, symmetric with respect to two vertical axes in order to implement the orientifold one needs, and containing the hexagonal cluster of our interest, together with a few additional technical constraints carefully discussed in \cite{Argurio:2020npm}. The hexagonal cluster can be inscribed in a disk; let us draw its ZZPs as in the middle of \Cref{Fig:hexagonalcluster} and keep only the `in' and `out' insertions on the boundary together with the pairing, as displayed on the right of \Cref{Fig:hexagonalcluster}. By construction the hexagonal cluster of interest can be obtained from this data.

\begin{figure}[h!]
	\centering
	\begin{subfigure}{0.23\textwidth}
		\centering
		\includegraphics[width=0.9\linewidth]{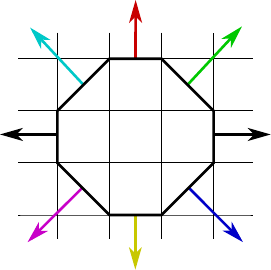} 
	\end{subfigure}
	\begin{subfigure}{0.23\textwidth}
		\centering
		\includegraphics[width=0.8\linewidth]{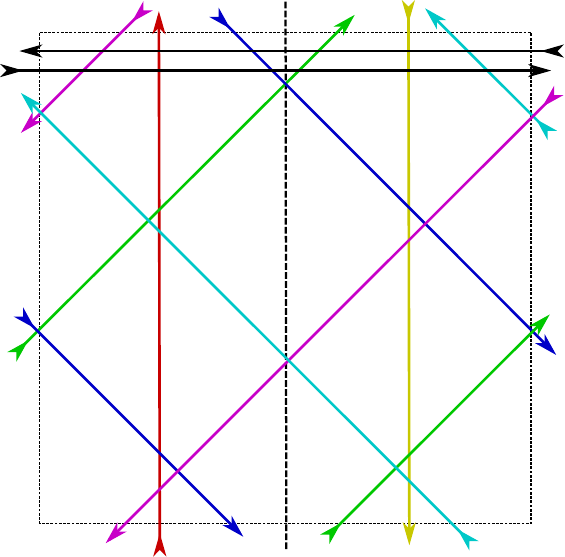} 
	\end{subfigure}
	\begin{subfigure}{0.23\textwidth}
		\centering
		\includegraphics[width=0.9\linewidth]{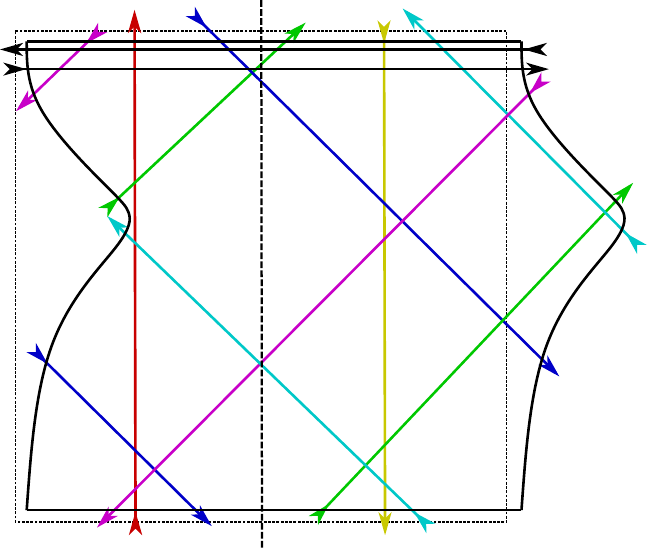} 
	\end{subfigure}
	\begin{subfigure}{0.23\textwidth}
		\centering
		\includegraphics[width=0.9\linewidth]{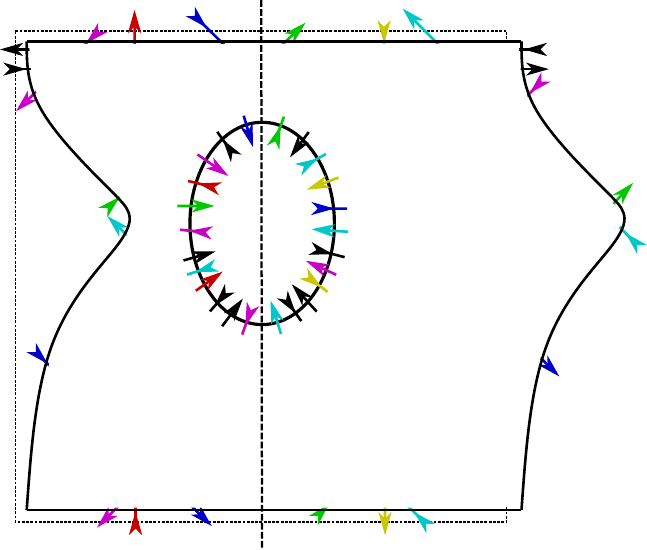} 
	\end{subfigure}
	\caption{From a polygon to a consistent dimer model through triple crossing diagrams.}
	\label{Fig:octagon}
\end{figure}

Our goal is to embed this circle (with in/out points and pairing) inside a consistent dimer model with the constraints listed above. One can show that a polygon that can possibly work must have at least $8$ sides; the octagon shown on the left of \Cref{Fig:octagon} is a parsimonious choice. One places straight lines on a fundamental cell as before, in a symmetric way with respect to the dashed vertical axis. Deforming the boundary of the fundamental cell so that endpoints of strands are distinct, one only keeps the in/out points and inserts the circle corresponding to the hexagonal cluster on the symmetry axis in the middle of the cell. The different steps of this procedure are displayed in \Cref{Fig:octagon}.

\begin{figure}[h!]
	\centering
	\begin{subfigure}{0.23\textwidth}
		\centering
		\includegraphics[width=0.9\linewidth]{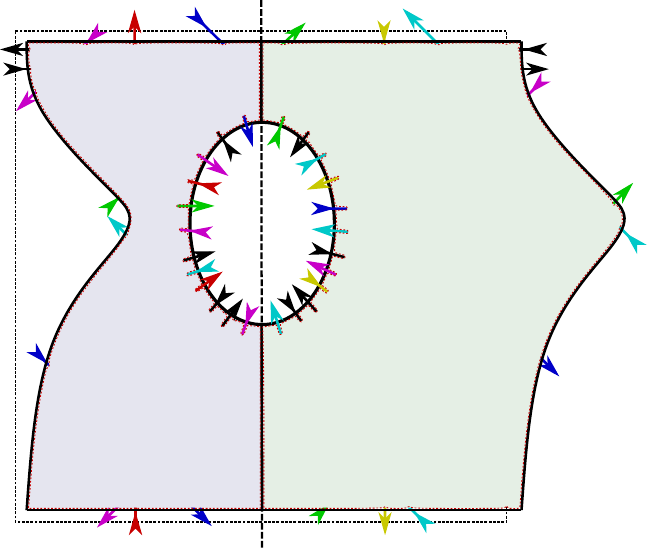} 
	\end{subfigure}
	\begin{subfigure}{0.23\textwidth}
		\centering
		\includegraphics[width=0.9\linewidth]{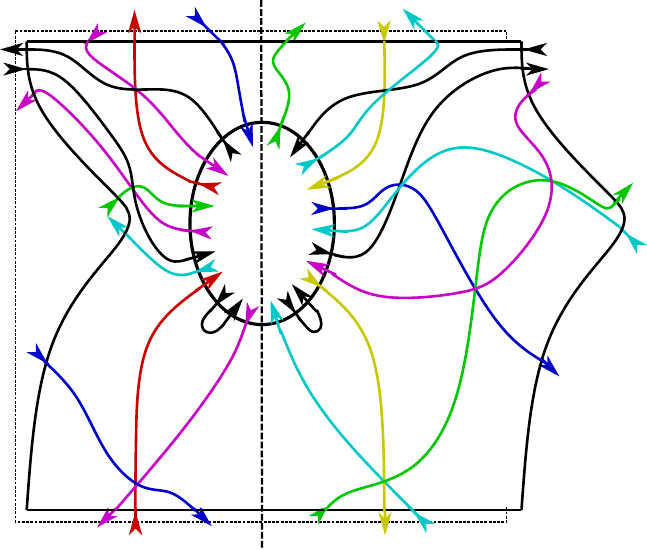} 
	\end{subfigure}
	\begin{subfigure}{0.23\textwidth}
		\centering
		\includegraphics[width=0.9\linewidth]{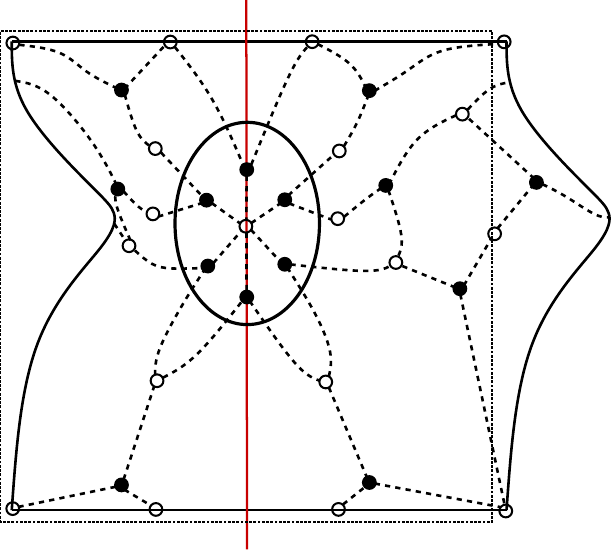} 
	\end{subfigure}
	\begin{subfigure}{0.23\textwidth}
		\centering
		\includegraphics[width=0.9\linewidth]{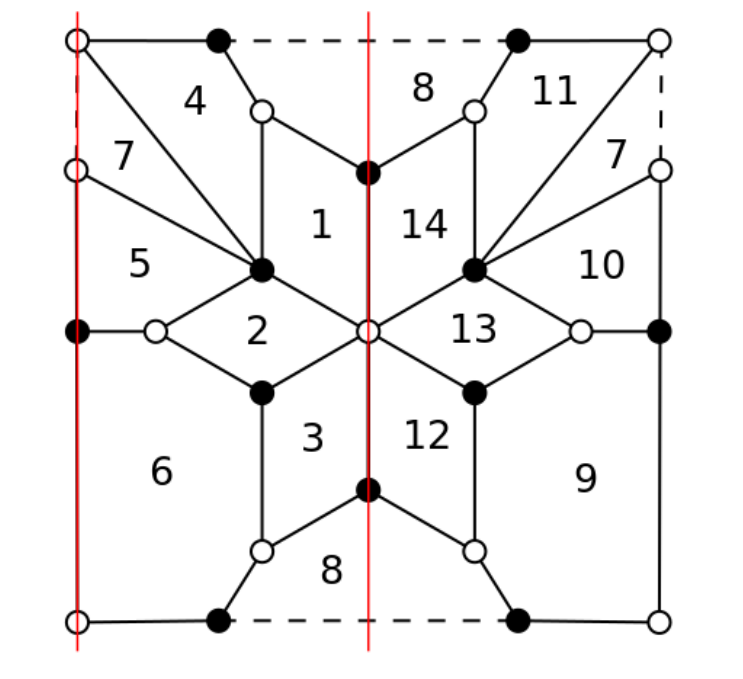} 
	\end{subfigure}
	\caption{The construction of the octagon dimer model.}
	\label{Fig:octagondimer}
\end{figure}

\vspace{-0.25cm}

The bluish and greenish parts of the deformed fundamental cell on the left of \Cref{Fig:octagondimer} are topological disks with boundary data and pairing satisfying Thurston's conditions, hence one can construct a (symmetric) triple crossing diagram realizing the boundary data, and run the algorithm as before to obtain a dimer model satisfying all the constraints listed above. It is presented under a fancy guise on the right of \Cref{Fig:octagondimer}. The physics of this Octagon dimer model is discussed in \cite{Argurio:2020dkg}.

\section{Symmetries in dimer models} 

\begin{figure}[h!]
	\centering
	\begin{subfigure}{0.4\textwidth}
		\centering
		\includegraphics[width=0.6\linewidth]{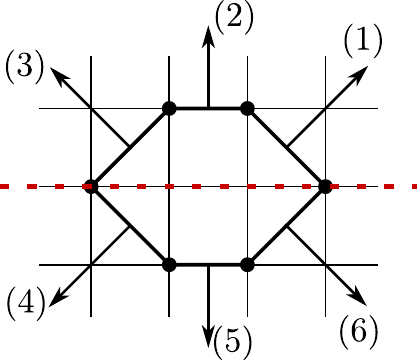} 
	\end{subfigure}
	\begin{subfigure}{0.4\textwidth}
		\centering
		\includegraphics[width=0.45\linewidth]{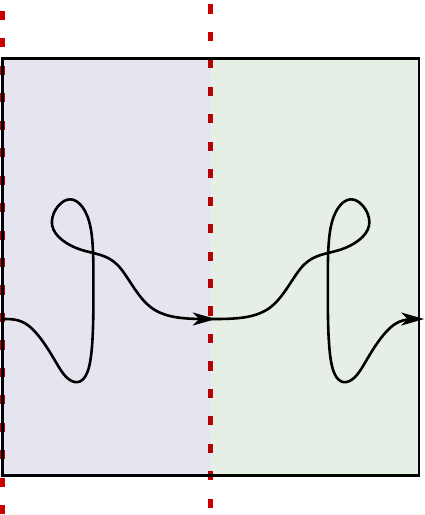} 
	\end{subfigure}
	\caption{From a polygon to a consistent dimer model through triple crossing diagrams.}
	\label{Fig:horizontaldp3}
\end{figure}

Triple point diagrams are very likely to be useful in order to study general aspects of the inverse algorithm and its geometric counterparts. Let us consider a specific example for concreteness. In \cite{Retolaza:2016alb} it is shown that if a dimer model is symmetric with respect to a vertical axis mapping vertices to vertices of the same color, then the corresponding toric diagram must be symmetric with respect to a horizontal axis\footnote{This follows by considering how the symmetry acts on ZZPs in the symmetric dimer model.}. One can wonder about the reciprocal: given a toric diagram symmetric with respect to a horizontal axis, is there always a corresponding dimer model symmetric with respect to a vertical axis? One can argue that this is false by considering the toric diagram on the left of \Cref{Fig:horizontaldp3}.

From the toric diagram it appears that no ZZP is mapped to itself under the reflection with respect to the dashed red line. Hence there must be edges on each symmetry axis in the dimer model, and subsequently it is always possible to choose a symmetric path as the black one displayed on the right of \Cref{Fig:horizontaldp3}, such that the total number of intersections (counted with signs) between this path and ZZPs in the left half of the cell is zero. Let $n_i\in\mathbb{Z}$, $i=1,\dots,6$ be the number of intersections (with signs) between the black path and the $i$-th ZZP in the left bluish half of the fundamental cell -- and remember that we have chosen the path so that $\sum n_i =0$. 

\textit{If the dimer model under consideration is symmetric with respect to the dashed red lines}, then the black path intersect (with signs) $-n_6$ times the ZZP $(1)$ on the right half of the cell since the ZZPs $(1)$ and $(6)$ are exchanged under the symmetry. Then, from the toric diagram one infers that $n_1-n_6=1$ since the winding of the black path is $(1,0)$ and hence as it loops around the torus it crosses once (with sign) the ZZPs $(1)$ whose winding is $(1,1)$. Similarly, $n_2-n_5=1$ and $n_3-n_4=1$. Hence $\sum n_i=2(n_4+n_5+n_6)+3$, which can not be zero since the $n_i$'s are integers, and we are led to the conclusion that there can not be any symmetric dimer model corresponding to this toric diagram and this symmetry. The example of above can be generalized straightforwardly to many other toric diagrams, and answers negatively the question of whether there always exist symmetric dimer models when the toric diagram is compatible with this symmetry. Thus one presumably needs intrinsically non-perturbative sectors \cite{Garcia-Etxebarria:2015hua} to describe the corresponding orientifolds, if any.

As one wonders about the existence of constrained dimer models, ZZPs-based arguments provide in general merely \textit{necessary} criteria -- such as the one we just derived, whereas the existence theorems for triple point diagrams let one hope for \textit{sufficient} conditions. The knowledge of these would certainly deepen our understanding of orientifolds of affine toric CY3 singularities.

\begin{center}
	$\star$ \hspace{0.05cm} $\star$ \hspace{0.05cm} $\star$
\end{center}

\vspace{0.15cm}

It is a pleasure to thank Riccardo Argurio, Matteo Bertolini, Sebasti\'an Franco, Eduardo Garc\'ia-Valdecasas, Shani Meynet and Antoine Pasternak for countless inputs and illuminating discussions, as well as comments on the manuscript. I would also like to thank Mo-Lin Ge, Cheng-Ming Bai, Yang-Hui He, Hong-Qin Li, Jiakang Bao, Edward Hirst and Suvajit Majumder for organizing this very nice symposium.

\footnotesize
\bibliographystyle{ieeetr}
\bibliography{ref}

\end{document}